# Nanolasers grown on silicon

Roger Chen, Thai-Truong D. Tran, Kar Wei Ng, Wai Son Ko, Linus C. Chuang, Forrest G. Sedgwick, Connie Chang-Hasnain*

*Department of Electrical Engineering and Computer Sciences and Applied Science and Technology Group, University of California at Berkeley, Berkeley, California 94720, USA*

* e-mail: cch@eecs.berkeley.edu

**Integration of optical interconnects with silicon-based electronics can address the growing limitations facing chip-scale data transport as microprocessors become progressively faster. However, material lattice mismatch and incompatible growth temperatures have fundamentally limited monolithic integration of lasers onto silicon substrates until now. Here, we use a novel growth scheme to overcome this roadblock and directly grow on-chip InGaAs nanopillar lasers, demonstrating the potency of bottom-up nano-optoelectronic integration. Unique helically-propagating cavity modes are employed to strongly confine light within subwavelength nanopillars despite low refractive index contrast between InGaAs and silicon. These modes thereby provide an avenue for engineering on-chip nanophotonic devices such as lasers. Nanopillar lasers are as-grown on silicon, offer tiny footprints and scalability, and are thereby particularly suited to high-density optoelectronics. They may ultimately form the basis of the missing monolithic light sources needed to bridge the existing gap between photonic and electronic circuits.**

Since the first laser demonstrated that stimulated emission processes in an optical medium can implement a powerful, coherent light source[1], the field of photonics has witnessed an explosion of applications in telecommunications, lighting, displays, medicine, and optical physics amongst others. Integration of photonic and electronic devices to leverage the advantages of both has subsequently attracted great interest. In particular, integration of optical interconnects onto silicon (Si) chips has become critical as ongoing miniaturization of Si logic elements has incurred a bottleneck in inter- and intra-chip communications[2,3]. Efforts towards creating on-chip light sources for optical interconnects have included engineering silicon and germanium for optical gain[4-6] and stimulated Raman scattering[7-9]. Concurrently, III-V lasers have been heterogeneously bonded onto silicon substrates[10-12]. However, numerous challenges face these approaches. Wafer bonding have low yields because of a stringent surface flatness requirement down to the atomic scale, while group IV emitters must overcome an indirect band gap that offers exceedingly inefficient radiation. Monolithic growth of high-performance III-V lasers on silicon thereby remains a "holy grail" for cost-effective, massively scalable, and streamlined fabrication of on-chip light sources.

The fundamental roadblock facing monolithic integration up to now has been a large mismatch of lattice constants and thermal expansion coefficients between III-V materials and silicon. Furthermore, growth of high-quality III-V materials traditionally occurs at high temperatures that are incompatible with silicon electronics. In order to overcome these barriers, we recently demonstrated growth of single crystal GaAs nanoneedles on silicon at 400 °C under conditions compatible with complementary metal-oxide-semiconductor (CMOS) technology[13,14]. Based on this template, we now develop InGaAs/GaAs heterostructure nanopillar lasers that are monolithically integrated onto Si after a single growth step. Compatible with silicon technology, nanopillar lasers can potentially leverage the capabilities of today's massive silicon infrastructure to facilitate an ongoing paradigm shift in modern computing architecture towards optoelectronic circuitry. In addition, we show a novel helically-propagating mode cavity that offers a unique feedback mechanism to enable on-chip laser oscillation. Whereas traditional Fabry-Perot (FP) modes are inhibited by the interface between InGaAs and Si, helical modes can strongly localize light within nanopillars of even subwavelength dimensions without lossy plasmonic or metal-optic effects. These modes thereby exemplify how novel optical phenomena in nanostructures can be used for future on-chip nanophotonic devices.

The nanopillar-based laser is schematically depicted in Figure 1A. Its cross-section reveals inner core-shell layers, while the inset offers a top view of the laser. The core contains the InGaAs active region while the GaAs shell provides surface passivation. Most nanopillars have a slight 5° taper between opposite sidewall facets. The height and diameter of the nanopillars scale easily and controllably with growth time despite the large lattice mismatch. Typical nanopillars possess tiny footprints of only ~0.34 $\mu m^2$, enabling them to realize high-density silicon-based optoelectronics. Nanopillars display extremely well-faceted geometry as shown by the scanning electron microscope (SEM) image in Fig. 1B. The top-view SEM image in Fig. 1C meanwhile shows the hexagonal cross-section of the nanopillar, which results from its unique single crystal wurtzite structure[15]. As we will later show, the as-grown nanopillar structure provides a natural optical cavity supporting unique resonances that have not been observed before to the best of our knowledge. As such, nanopillars do not require additional top-down processing to form on-chip optical cavities. Instead, they provide a viable bottom-up approach towards integrating light sources and resonators onto a silicon chip.

Importantly, nanopillars possess several critical advantages for optoelectronic integration onto silicon. They grow at a low temperature of 400 °C, which is drastically lower than typical III-V growth temperatures by 200-300 °C and is compatible with CMOS devices. Heterostructure nanopillar growth occurs spontaneously on a silicon substrate by metal-organic chemical vapor deposition (MOCVD) and is catalyst-free; thus, we avoid incorporating metal particles that are poisonous to Si CMOS devices. Additionally, core-shell layering allows nanopillar laser diode structures to be grown on silicon. Detailed procedures are described in Supplementary Information. Low temperature and catalyst-free growth allows nanopillars to be monolithically integrated with silicon electronics without compromising highly-developed Si transistor technology and CMOS process flows.

Room-temperature laser oscillation is achieved by optical excitation using a modelocked Ti:sapphire laser (see Supplementary Information). At low pump levels, nanopillars emit broad spontaneous emission. With increasing excitation intensity, a cavity mode emerges and ultimately laser oscillation is seen at ~950 nm, as shown in Fig. 2A. The inset shows the nanolaser light output power and emission spectral linewidth as functions of excitation (the former is referred to as the 'light-light' or 'L-L curve'). Clear threshold behavior is observed in the L-L curve along with prominent clamping of the laser linewidth. We observe a classic signature of lasing from near-field images of nanopillar lasers. When pumped below threshold, a nanopillar only outputs a spot of spontaneous emission as shown in Fig. 2B. Above threshold, clear speckle patterns due to interference of highly-coherent light become visible in Fig. 2C. Most lasers characterized show single mode operation, though as many as three lasing modes have been seen from a single nanopillar. We note that growing a sufficiently thick GaAs cap (~90 nm) is key for suppressing surface velocity to enable room temperature operation. Lasing on silicon at room temperature is a critical achievement demonstrating the potential of nanopillar lasers for implementing practical optoelectronics beyond the research lab.

We perform additional experimental studies at 4 K to further characterize the nanopillar laser and extract various laser parameters. Fig. 2D shows the L-L curve and the corresponding spectra below and above lasing threshold for a typical laser: a clear threshold is seen at a very low excitation intensity of ~22 µJ/cm$^2$. By constructing an analytical gain model using material parameters from literature[16], we analyze the nanopillar's spontaneous emission spectrum and establish a correlation between experimental pump levels and carrier density. Applying the gain model with classical rate equation analysis[17,18], we fit the L-L curve in Fig. 2D to reveal a threshold gain and carrier density of ~1,400 $cm^{-1}$ and ~$1\times10^{18}$ $cm^{-3}$, respectively (see Supplementary Information). We thereby estimate a cavity quality ($Q$) factor of ~206 and a spontaneous emission factor of $β$~0.01, where $β$ measures the fraction of spontaneous emission coupled into the cavity mode of interest. This value is reasonable considering the laser's small volume of $6\times10^{-13}$ $cm^3$ [19]. We additionally measure a strong background suppression ratio of ~17-20 dB from typical lasing spectra.

Despite minimal index contrast between the InGaAs nanopillar ($n_r$~3.7) and the Si substrate it resides on ($n_r$~3.6), sufficient optical feedback or $Q$ is attained for laser oscillation in our novel structures. Even non-negligible absorption by silicon at nanopillar laser wavelengths is overcome. Intuitively, helically-propagating modes can have nearly total internal reflection at the nanopillar-Si interface because their wavevectors strike that boundary with grazing incidence at extremely shallow angles. Fig. 3A illustrates this novel concept using a ray optics interpretation. Alternatively, the interface between substrate and nanopillar can be thought of as introducing a strong cutoff for helically-propagating modes. The strong feedback of helical modes is further confirmed by finite-difference time-domain (FDTD) calculations of untapered nanopillars. These helical nanopillar modes possess strong azimuthal components, which result in transverse field profiles similar to those previously reported for hexagonal whispering gallery (WG) modes[20-22] as shown in Fig. 3B. However, unlike traditional WG modes, helical nanopillar resonances have net propagation in the axial direction. This can be seen from the first and third order axial standing wave patterns shown in Figs. 3C and 3D, respectively. We can therefore define two mode numbers to describe these helically-propagating modes. An azimuthal mode number m describes the transverse field pattern just as they do for WG modes, while an axial mode number n describes the order of the axial standing wave like for FP resonances. To be complete, a third mode number should account for higher-order radial modes, but we find that the subwavelength transverse dimensions of nanopillars limit access to only the lowest-order radial mode. Both transverse electric (TE) as well as transverse magnetic (TM) polarizations exist. As might be expected, modes with distinct mode numbers resonate at distinct wavelengths. Both nanopillar radius and length determine cavity resonances, further substantiating the helical nature of nanopillar cavity modes. Traditional WG resonances are meanwhile relatively insensitive to the axial dimension. In our description, they can actually be interpreted as the n=1 subset of helically-propagating modes. We find that most nanopillars lase between 890-930 nm when the nanopillar radius is varied between 270 nm to 340 nm. This range corresponds to TM m=5, 6 and TE m=4, 5 resonances, which are among the lowest-order azimuthal modes reported to have achieved laser oscillation. Additional discussion and details of FDTD calculations and results are presented in Supplementary Information.

The most important consequence of a helically propagating mode is that it allows high reflectivity at a dielectric interface with low index contrast. From Figs. 3C and 3D, it is evident that the cavity modes are remarkably well-confined in the active material despite the significant interface with silicon. Such strong confinement enables our resonators to lase without having to fabricate the pedestals that WG microdisk lasers typically require to preserve vertical optical confinement. In fact, the confinement is strong enough that even nanopillars with subwavelength dimensions on all sides achieve laser oscillation. An SEM image of such a subwavelength laser is shown in Fig. 3E. The physical volume of that laser is only V~$0.2\lambda_o^3$, where $\lambda_o$ is the lasing wavelength in air. Subwavelength lasers have meanwhile been successfully implemented on other substrates by use of plasmonics[23-25]. In our case, a purely semiconductor cavity mode alone provides enough confinement for subwavelength lasing without metal-optic effects[26,27]. Avoiding metal losses is a critical advantage that should allow high-efficiency laser operation.

Engineering these helical modes by controlling nanopillar dimensions, we can select the mode number to be used for laser oscillation and control the wavelength of nanopillar laser emission. Since nanopillar cavities scale with growth time from the nanoscale throughout the microscale without critical dimensions, we can easily grow nanopillars to resonate at any wavelength of choice. Experimentally, we verify that nanopillars support n>1 resonances by directly imaging emission from nanopillars of different lengths as presented in Figs. 3F-H with corresponding SEM images. Axial standing wave patterns can be clearly discerned, confirming net propagation of nanopillar cavity modes along the nanopillar axis. As nanopillar length decreases, higher-order axial modes are cut off such that fewer axial maxima can be observed. Nanopillars are held below threshold for these images to prevent speckle interference from obscuring the mode pattern. Traditional FP modes cannot account for the patterns seen since light leakage through the InGaAs-silicon interface is far too severe for such modes. Instead, we attribute the axial maxima to helically-propagating modes of the nanopillar, which simulations show to be strongly supported despite even oblique junctions of low index contrast between nanopillars and substrate.

By varying the indium composition of nanopillars during growth, we can spectrally match the semiconductor gain to designed cavity resonances to maximize spontaneous emission coupling into the cavity modes (see Supplementary Information)[28-30]. In this current study, we vary indium composition from 12-20% to achieve wavelength control of on-chip nanolasers over a ~50 nm range as shown in Fig. 4. The flexibility of wavelength control may allow nanopillars to fulfill a myriad of laser applications throughout the near-infrared. In the future, we expect to push nanopillar emission beyond silicon transparency, which would improve cavity $Q$ by removing Si absorption loss. We remark that our experimental $Q$ falls well short of theoretical estimates (up to ~4,300), likely because tapering in the nanopillar perturbs the mode. Further engineering of the nanopillar geometry thus offers another means of future improvement. The fact that we grow laser cavities using a bottom-up approach rather than top-down etching processes raises a couple additional points of interest. First, single crystal growth provides facets that are far smoother than etching can allow, thereby minimizing scattering loss. Additionally, the hexagonal structure of nanopillars is dictated by their hexagonal wurtzite crystal lattice. Thus, as-grown nanopillars have unprecedented symmetry, which prevents polarization splitting of degenerate laser modes often seen in microdisks due to ellipticity. Instead, truly single mode laser oscillation can be achieved.

In this letter, we demonstrate the first room-temperature III-V nanolaser grown on Si with subwavelength volume. In doing so, we show that lattice mismatch is not a fundamental limit for monolithic integration of III-V photonic devices onto silicon as is generally perceived. A bottom-up approach can provide an effective way of integrating nanophotonics with nanoelectronics. The ability to heavily populate an optically-deficient silicon substrate with efficient III-V lasers has far-reaching implications for silicon photonics, particularly with regard to optical interconnects. A new class of helically-propagating modes implement our unique on-chip optical cavities, showing that optical phenomena in all-semiconductor nanostructures can be leveraged for practical on-chip nanophotonic devices. These modes may be leveraged for new designs of optical components such as photodetectors, modulators and solar cells. Future electrical operation of p-i-n nanopillar lasers on silicon meanwhile promises for a powerful marriage between photonic and electronic circuits.

**Methods**

**Growth.** To grow InGaAs nanopillars, a silicon substrate is first cleaned with acetone, methanol, and water for 3 minutes at each step. Afterwards, the substrate is deoxidized by buffered oxide etch for 3 minutes, and the surface is mechanically roughened. Growth is subsequently carried out in an EMCORE D75 MOCVD reactor. Tertiarybutylarsine (TBA) is introduced to the reactor at temperatures higher than 200 °C. Before growth, in-situ annealing at 600 °C is performed for 3 minutes. After annealing, the temperature is reduced to the growth temperature, which is 400 °C in this work, in 3 minutes, followed by 2 minutes of temperature stabilization. Triethylgallium (TEGa) and Trimethylindium (TMIn) are then introduced to the reactor to begin the 60-min InGaAs core growth. TMIn mole fractions are kept constant at $9.86 \times 10^{-7}$, $1.38 \times 10^{-6}$, and $1.73 \times 10^{-6}$ for 12%, 15% and 20% indium compositions, respectively. The TEGa mole fraction is held at $1.12 \times 10^{-5}$. All sources use a 12 liter/min hydrogen carrier gas flow. The TBA mole fraction is $5.42 \times 10^{-4}$; hence, the V/III ratio is ~43. A GaAs shell is then grown around the InGaAs core with the same TEGa and TBA mole fractions used for core growth, with a V/III ratio of 48. Room temperature lasers employ 90 nm shells with 20% indium cores, while low temperature lasers typically have 30 nm shells and 15% indium cores. Nanopillar growth is vertically aligned to the (111) silicon substrate and anisotropic with faster growth rates along the [0001] wurtzite c-axis. Nanopillar dimensions are linearly scalable with time with no critical dimensions observed.

**Lasing experiments.** To achieve nanopillar laser oscillation, 120 fs pump pulses from a modelocked Ti-sapphire laser (Coherent Mira, $\lambda_{pump}$=750 nm, repetition rate 76 MHz) are delivered by a 100x 0.7 numerical aperture objective (Mitutoyo NIR HR) to the sample at a temperature held between 4 K and 293 K by a continuous-flow liquid helium cryostat (Oxford Instruments Hi-Res II). The pump spot size is slightly defocused to ~6 µm in diameter. Nanopillar emission is collected by the same objective and relayed to a spectrometer and $LN_2$-cooled silicon charge-coupled device (Princeton Instruments SP2750 and Spec-10). Filters are used in all experiments to prevent pump light from reaching any detectors or cameras.

# References

1. Maiman, T. H. Stimulated optical radiation in ruby. *Nature* **187**, 493–494 (1960).
2. Miller, D. A. B. Device Requirements for Optical Interconnects to Silicon Chips. *Proc. IEEE* **97**, 1166-1185 (2009).
3. Chen, G. *et al*. Predictions of CMOS compatible on-chip optical interconnect. *Integration, the VLSI Journal* **40**, 434-446 (2007).
4. Cloutier, S. G., Kossyrev, P. A. & Xu, J. Optical gain and stimulated emission in periodic nanopatterned crystalline silicon. *Nature Mater.* **4**, 887-891 (2005).
5. Pavesi, L., Dal Negro, L., Mazzoleni, C., Franzo, G. & Priolo, F. Optical gain in silicon nanocrystals. *Nature* **408**, 440-444 (2000).
6. Liu, J., Sun, X., Camacho-Aguilera, R., Kimerling, L. C. & Michel, J. Ge-on-Si laser operating at room temperature. *Opt. Lett.* **35**, 679-681 (2010).
7. Rong, H. *et al*. An all-silicon Raman laser. *Nature* **433**, 292-294 (2005).
8. Rong, H. *et al*. A continuous-wave Raman silicon laser. *Nature* **433**, 725-728 (2005).
9. Boyraz, O. & Jalali, B. Demonstration of a silicon Raman laser. *Opt. Express* **12**, 5269–5273 (2004).
10. Fang, A.W. *et al*. Electrically pumped hybrid AlGaInAs-silicon evanescent laser. *Opt. Express* **14**, 9203-9210 (2006).
11. Van Campenhout, J. *et al*. Electrically pumped InP-based microdisk lasers integrated with a nanophotonic silicon-on-insulator waveguide circuit. *Opt. Express* **15**, 6744–6749 (2007).
12. Lo, Y. H., Bhat, R., Hwang, D. M., Chua, C. & Lin, C.-H. Semiconductor lasers on Si substrates using the technology of bonding by atomic rearrangement. *Appl. Phys. Lett.* **62**, 1038 (1993).
13. Moewe, M., Chuang, L. C., Crankshaw, S., Chase, C. & Chang-Hasnain, C. Atomically sharp catalyst-free wurtzite GaAs/AlGaAs nanoneedles grown on silicon. *Appl. Phys. Lett.* **93**, 023116 (2008).
14. Moewe, M., Chuang, L. C., Crankshaw, S., Ng, K. W. & Chang-Hasnain, C. Core-shell InGaAs/GaAs quantum well nanoneedles grown on silicon with silicon-transparent emission. *Opt. Express* **17**, 7831-7836 (2009).
15. Chen, R. *et al*. Second-harmonic generation from a single wurtzite GaAs nanoneedle. *Appl. Phys. Lett.* **96**, 051110 (2010).
16. Chuang, S. L. *Physics of Optoelectronic Devices* 337-393 (Wiley, New York, 1995).
17. Coldren, L. A. & Corzine, S. W. *Diode Lasers and Photonic Integrated Circuits* 185-261 (Wiley, New York, 1995).
18. Baba, T. Photonic crystals and microdisk cavities based on GaInAsP-InP system. *IEEE J. of Sel. Top. in Quant. Electron.* **3**, 808–830 (1997).
19. Bjork, G. & Yamamoto, Y. Analysis of semiconductor microcavity lasers using rate equations. *IEEE J. Quant. Electron.* **27**, 2386-2396 (1991).
20. McCall, S. L., Levi, A. F.J., Slusher, R. E., Pearton, S. J. & Logan, R. A. Whispering-gallery mode microdisk lasers. *Appl. Phys. Lett.* **60**, 289-291 (1992).
21. Nobis, T. & Grundmann, M. Low-order optical whispering-gallery modes in hexagonal nanocavities. *Phys. Rev. A* **72**, 063806 (2005).
22. Wiersig, J. Hexagonal dielectric resonators and microcrystal lasers. *Phys. Rev. A* **67**, 023807 (2003).
23. Hill, M. T. *et al*. Lasing in metallic-coated nanocavities. *Nature Photon.* **1**, 589-594 (2007).

24. Oulton, R. F. *et al*. Plasmon lasers at deep subwavelength scale. *Nature* **461**, 629-632 (2009).
25. Hill, M. T. *et al*. Lasing in metal-insulator-metal sub-wavelength plasmonic waveguides. *Opt. Express* **17**, 11107–11112 (2009).
26. Nezhad, M. P. *et al*. Room-temperature subwavelength metallo-dielectric lasers. *Nature Photon.* **4**, 395-399 (2010).
27. Yu, K., Lakhani, A. & Wu, M. C. Subwavelength metal-optic semiconductor nanopatch lasers. *Opt. Express* **18**, 8790-8799 (2010).
28. Kuykendall, T., Ulrich, P., Aloni, S. & Yang, P. Complete composition tunability of InGaN nanowires using a combinatorial approach. *Nature Mater.* **6**, 951-956 (2007).
29. Qian, F. *et al*. Multi-quantum-well nanowire heterostructures for wavelength-controlled lasers. *Nature Mater.* **7**, 701-706 (2008).
30. Pan, A. *et al*. Continuous alloy-composition spatial Grading and superbroad wavelength-tunable nanowire lasers on a single chip. *Nano Lett.* **9**, 784–788 (2009).

**Supplementary Information** accompanies the paper on **www.nature.com/nature**.

**Acknowledgements** This work was supported by DARPA UPR Award HR0011-04-1-0040, MARCO IFC, and a DoD National Security Science and Engineering Faculty Fellowship via Naval Post Graduate School N00244-09-1-0013. R.C. acknowledges support from a National Defense Science and Engineering Graduate Fellowship. The authors acknowledge Professor Cun-Zheng Ning and Dr. Erwin K. Lau for fruitful discussions.

**Author Contributions** R.C., T.-T.D.T, K.W.N. and C.C.H. designed the experiments. R.C. and T.-T.D.T. performed optical measurements. K.W.N., W.S.K. and L.C.C. developed and performed material growth. K.W.N. and W.S.K. performed SEM measurements. R.C., T.-T.D.T. and F.G.S. studied and simulated the optical mode. R.C. performed gain modeling and rate equation analysis. R.C. and C.C.H. composed the manuscript. C.C.H. guided the overall project.

**Author Information** Reprints and permissions information is available at npg.nature.com/reprintsandpermissions. Correspondence and requests for materials should be addressed to C.C.H. (cch@eecs.berkeley.edu).

**Figure 1 InGaAs/GaAs heterostructure nanopillar lasers monolithically grown on silicon. a,** A schematic is shown for a nanopillar laser monolithically integrated onto silicon, illustrating its InGaAs core and GaAs shell. The higher band gap GaAs shell protects carriers from nonradiative surface recombination, which is critical for room temperature lasing. The inset shows a top-view schematic. **b,** An SEM image reveals the well-faceted geometry of the nanopillar optical cavity. This resonator structure forms naturally during growth, allowing lasers to be directly grown on silicon without additional processing. Nanopillar dimensions scale with growth time, enabling growth of effectively bulk high-quality III-V material on silicon. No critical dimensions have been observed for this novel lattice-mismatched growth mechanism. **c,** A top-view SEM image highlights the nanopillar's hexagonal wurtzite crystal structure. This hexagonal symmetry results in whispering gallery-like effects that we discuss later.

**Figure 2 On-chip nanopillar laser oscillation. a,** Room temperature nanopillar emission is shown below (blue) and above (red) threshold. Below threshold, broad spontaneous emission is seen. The spectrum below threshold has been magnified 200x for visibility. Above threshold, single mode laser oscillation is observed with a narrow lasing peak dominating nanopillar emission. Room temperature lasing attests to the high optical quality of nanopillars on silicon and is critical for practical implementation of optical interconnects. The inset shows the L-L curve and linewidth clamping of the laser, revealing a threshold at ~93 µJ/cm$^2$. **b,** Camera images of nanopillar emission below threshold show only a spot of spontaneous emission. **c,** Upon lasing, strong speckle patterns appear. Speckle results from the high degree of coherent emission and is a classic signature of laser oscillation. **d,** The L-L curve of a nanopillar laser operating at 4 K with a low threshold of ~22 µJ/cm$^2$ is analyzed to extract various laser characteristics.

Gain modeling is combined with traditional rate equation analysis to fit the L-L curve and extract estimated values of $Q$~206 and $\beta$~0.01. Details of the analysis are further discussed in the main text and Supplementary Information. The inset shows spectra of the nanopillar laser both below (blue) and above (red) threshold. The below threshold spectrum has been magnified 20x for visibility. The laser peak clearly dominates above threshold, achieving a 17 dB background suppression ratio.

**Figure 3 Helically-propagating modes for optical feedback of on-chip nanolasers. a,** Sufficient reflection from the low-index contrast interface between nanopillar and silicon can be achieved if light strikes that interface at glancing angles. As shown in the schematic, a helical ray path enables this, lending physical insight into the mechanism behind strong optical confinement of on-chip nanopillar cavities. The nanopillar-Si interface can also be interpreted as introducing a cutoff for these helical modes. **b,** An FDTD-simulated field profile shows a hexagonal WG-like mode pattern (TM m=6) in the transverse plane, which arises from the strong azimuthal components of the nanopillar's helically-propagating modes. Since nanopillar cavity modes have net propagation along the nanopillar axis, they build axial standing wave patterns akin to FP resonances. The first-order standing wave is shown in **c** while a higher-order one is shown in **d**. We define mode numbers m and n to describe the azimuthal and longitudinal field patterns of helical modes, respectively. Higher-order radial modes also exist for helical resonances, but our small dimensions limit us to the lowest-order radial mode only. Modes with different mode numbers have different resonances. For example, the modes shown in **c** and **d** resonate at $\lambda_1$=921 nm (n=1) and $\lambda_3$=916 nm (n=3), respectively, despite sharing the same transverse field profile (TM m=6) shown in **b**. Quasi-WG modes observed in traditional microdisks are equivalent to the n=1 subset of helically-propagating resonances. **e,** These helical modes can strongly confine light in small volumes even when the substrate offers little index contrast, enabling nanolasers that are subwavelength on all sides. An SEM image shows one of these subwavelength nanopillar lasers. **f-g,** Experimental images of nanopillar emission below threshold show standing wave patterns along the nanopillar axes, experimentally corroborating helical mode propagation. The imaged mode patterns are attributed to helical modes rather than Fabry-Perot modes because the existing interface between silicon and InGaAs prevents Fabry-Perot standing waves from forming. As shown, higher-order axial modes are cut off for shorter nanopillars, resulting in fewer axial emission maxima. SEM photos of the nanopillars imaged are also shown.

**Figure 4 Wavelength control of nanopillar lasers by composition variation.**
Nanopillar lasers can be implemented over a broad range of wavelengths by properly tuning indium composition and nanopillar dimensions. Nanopillars are grown to target dimensions designed to achieve resonances at specified wavelengths. Indium composition of the InGaAs core is concurrently adjusted so that nanopillar gain and emission spectrally match the resonant wavelength for laser oscillation. The spectra shown are offset for clarity. In this study, a change in indium composition from approximately 12-20% provides a 50 nm control in the output laser wavelength.

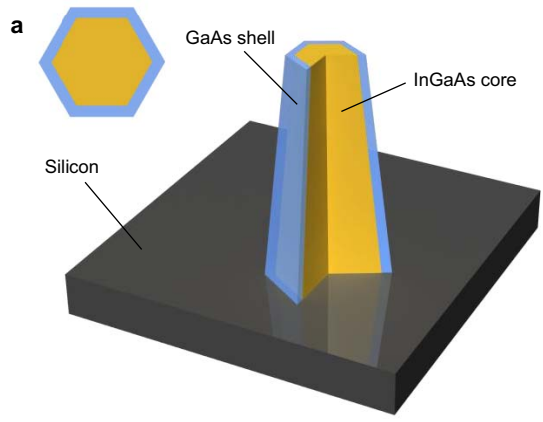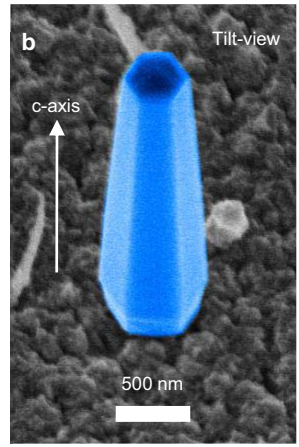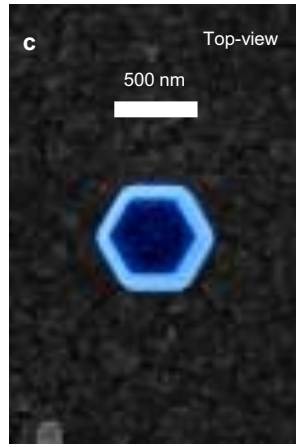

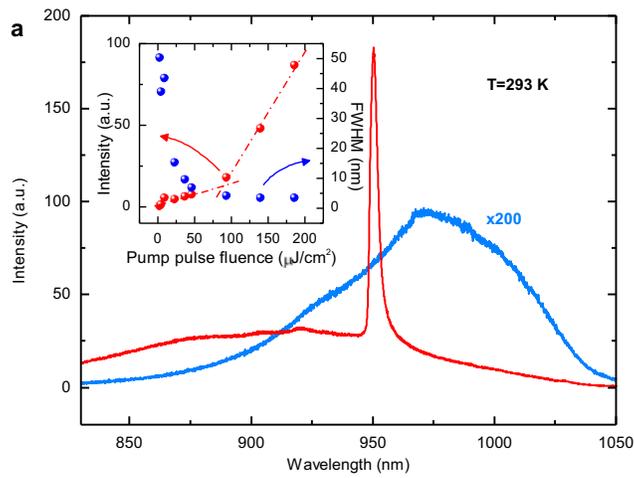

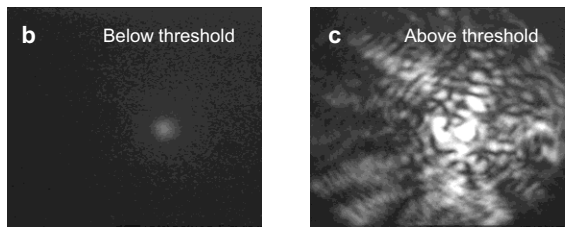

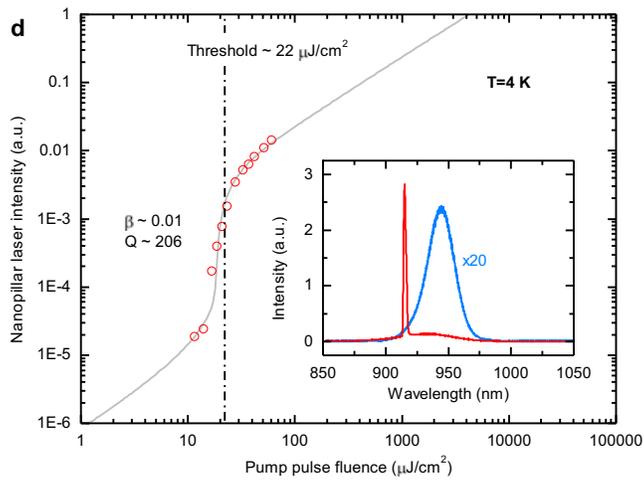

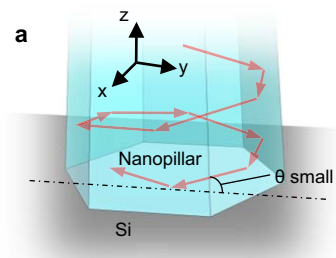
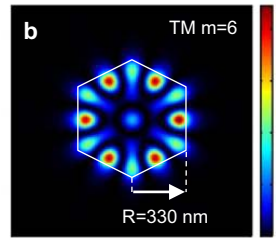
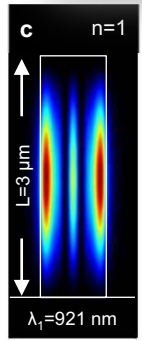
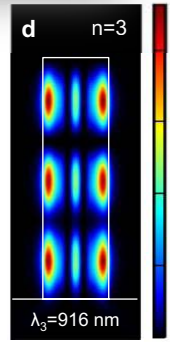
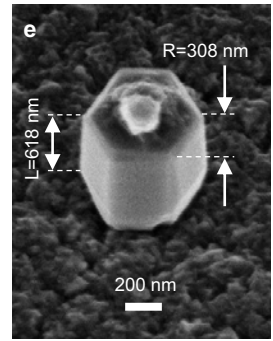
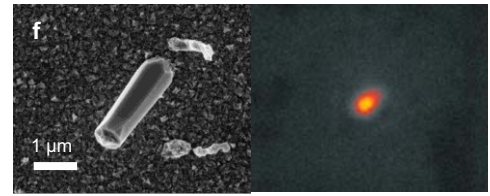
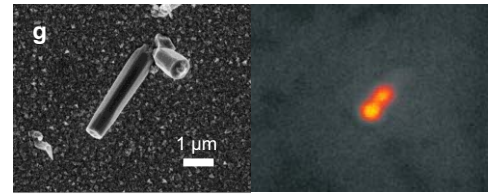
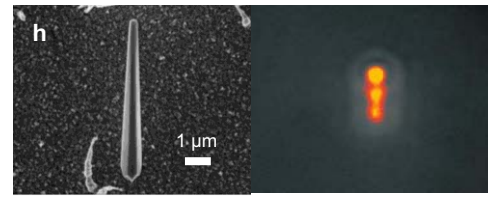

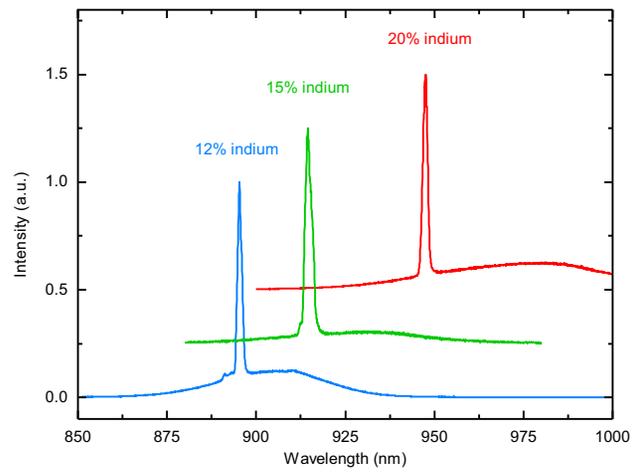